\documentclass[showpacs,preprintnumbers,preprint]{revtex4}

\usepackage{graphicx, epsfig}
\usepackage{bm}

\begin{document}

\preprint{SU-GP-02/7-2}
\preprint{SU-4252-766}
\title{Quintessential Baryogenesis}

\author{Antonio De Felice, Salah Nasri and Mark Trodden}

\affiliation{Department of Physics \\
Syracuse University \\
Syracuse, NY 13244-1130\\}

\date{\today}

\begin{abstract}
The simplest explanation for early time acceleration (inflation) and the late time
acceleration indicated by recent data is that they have a common origin. We investigate
another generic cosmological implication of this possiblity, that the baryon asymmetry
of the universe may be generated in such models. We identify several novel features of
baryogenesis in such a universe, in which a rolling scalar field is always part of the cosmological
energy budget. We also propose a concrete mechanism by which the baryon asymmetry of
the universe may be generated in this context. We analyze the generic properties of and constraints
on these cosmologies, and then demonstrate explicitly how a complete cosmology may develop in
some specific classes of models.
\end{abstract}

\pacs{98.80.Cq, 11.30.Er, 11.30.Fs, 12.60.Fr}

\maketitle

\section{Introduction}
\label{intro}
Rolling scalar fields are a mainstay of modern cosmology. This is perhaps best-illustrated by the 
inflationary paradigm~\cite{Guth:1980zm,Linde:1981mu,Albrecht:1982wi}, in which most implementations involve a scalar field rolling towards the
minimum of its potential in such a way that the potential energy of the field is the dominant
component of the energy density of the universe. There are, however, many other cosmological
instances in which scalar fields are invoked.

During the last few years a new consistent picture of the energy budget of the universe has emerged.
Large scale structure studies show that matter (both luminous and dark) contributes a fraction of
about 0.3 of the critical density, while the position of the first acoustic peak of the cosmic microwave
background power spectrum indicates that the total energy density is consistent with criticality. The
discrepancy between these two measurements may be reconciled by invoking a negative pressure 
component
which is termed {\it dark energy}. While there are a number of different observational tools to study
dark energy -- number counts of galaxies~\cite{Newman:1999cg}
and galaxy clusters~\cite{Haiman:2000bw} for example -- the most direct evidence 
to date comes from the light-curve measurements of intermediate redshift type IA 
supernovae~\cite{Riess:1998cb,Perlmutter:1998np}.
Consistency between these observations and others such as weak gravitational 
lensing~\cite{Huterer:2001yu} and
large scale structure surveys~\cite{Hu:1998tk} implies that the dark energy $X$ satisfy 
$\Omega_X \sim 0.7$ and that the equation of state be~\cite{Perlmutter:1999jt,Wang:1999fa}
\begin{equation}
w_X \equiv \frac{p_X}{\rho_X}\leq -0.6 \ ,
\end{equation}
leading to the acceleration of the universe.

It is of course possible that this mystery component is a cosmological constant $\Lambda$, for
which $w_{\Lambda}=-1$. However,
understanding the nature of such an unnaturally small $\Lambda$ is at least as difficult as undestanding
one that is zero. Alternatively, it has been suggested~\cite{Wetterich:fm}-\cite{Caldwell:1997ii}
that if the cosmological constant itself is zero,
the dark energy component could be due to the dynamics of a rolling scalar field, in a form of
late-universe inflation that has become known as {\it quintessence}. Although there are a number of
fine-tuning problems associated with this idea, it does provide a way to ensure the late-time
acceleration of the universe, albeit at the expense of introducing a second (after the inflaton)
cosmologically relevant rolling scalar field. While not addressing the cosmological constant 
problem itself, and suffering from fine-tuning, quintessence itself has the advantage of avoiding a future
horizon in space-time, and hence makes consistency with what is known about perturbative string
theory more likely.

It is natural to wonder whether the inflaton and the quintessence field might be one and the 
same~\cite{Spokoiny:1993kt}.
In fact, specific models for this have been 
proposed~\cite{Spokoiny:1993kt}-\cite{Dimopoulos:2001ix}. Clearly such models are attractive 
because we need only postulate a single rolling scalar, but may be problematic either theoretically or
phenomenologically.

In this paper we investigate how we may further limit the proliferation of rolling scalar fields
required in modern cosmology by studying how the scalar field responsible for late-time
acceleration of the universe might also solve another outstanding 
cosmological puzzle. Specifically we will be
interested in the role that such a field may play in the generation of the baryon asymmetry
of the universe. The spectacular success
of primordial nucleosynthesis requires that there exist an asymmetry
between baryons and antibaryons in the universe at temperatures lower than an MeV. This is 
quantified by the requirement
\begin{equation}
\label{BAUobserved}
4\times 10^{-10}\leq \eta \equiv \frac{n_B}{s} \leq 7\times 10^{-10} \ ,
\end{equation}
where $n_B\equiv n_b - n_{\bar b}$, with $n_{b({\bar b})}$ the number density of (anti)baryons and
$s$ is the entropy density. To generate such an asymmetry, the underlying particle physics
theory must satisfy three necessary conditions -- the Sakharov conditions~\cite{Sakharov:dj}. These
are baryon number $B$ violation, the violation of the discrete symmetries $C$ and $CP$ and a 
departure from thermal equilibrium, this last condition resulting from an application of the $CPT$ 
theorem. In this paper we are interested in how these conditions may be met within the context of
dark energy models.

The relationship between early-time acceleration -- inflation -- and baryogenesis has been 
explored
in some detail (for example see~\cite{Affleck:1984fy}-\cite{Nanopoulos:2001yu}).
Here we investigate the opposite regime, that the quintessence field may be associated with the
generation of the baryon asymmetry. Naturally, it would be 
particularly efficient if a single scalar field could be responsible for three fundamental phenomena
in cosmology -- inflation, baryogenesis and dark energy, and indeed we will show that baryogenesis
occurs quite generically in models in which a single scalar is responsible for the two periods
of cosmic acceleration.

The outline of this paper is as follows. In section~\ref{quintinf} we will review some details about
quintessence and explain how inflation and quintessence may be
unified by generalizing the quintessential inflation model of Peebles and Vilenkin~\cite{Peebles:1998qn}.
In section~\ref{qbg} we will describe how quintessence and quintessential inflation may naturally yield a 
baryon asymmetry
without the introduction of any new fields into the theory. We term this model {\it quintessential
baryogenesis}, borrowing the phrasing from Peebles and Vilenkin. 
This turns out to depend to some extent on the details of 
quintessential inflation. In section~\ref{sec:constraints} we will discuss experimental and astrophysical 
constraints on our models and comment on how we may test the physics involved. We offer our
comments and conclusions in the final section of the paper.

\section{Quintessence and Generalized Quintessential Inflation}
\label{quintinf}
As we have mentioned already, one approach to the dark energy problem is to assume that there is 
some as yet
unknown process that sets the vacuum cosmological constant of the universe to zero, but that there
exists a cosmologically-relevant scalar field in the universe, that has yet to reach its global minimum
and therefore contributes an effective vacuum energy to the total. This idea has been termed 
quintessence and we shall briefly review it here.

The Einstein equations in cosmology may be written as
\begin{equation}
\label{friedmann}
\left(\frac{{\dot a}}{a}\right)^2=\frac{8\pi}{3M_{\rm p}^2}\rho \ ,
\end{equation}
\begin{equation}
\label{acceleration}
\frac{{\ddot a}}{a}=-\frac{4\pi}{3M_{\rm p}^2}(\rho+3p) \ ,
\end{equation}
where we are using the Friedmann, Robertson-Walker (FRW) ansatz for the metric
\begin{equation}
\label{metric}
ds^2=-dt^2+a(t)^2\left[dr^2+r^2d\Omega_2^2\right] \ .
\end{equation}
Here the energy density $\rho$ and pressure $p$ for a real homogeneous scalar field $\phi$
are given by
\begin{equation}
\label{energydensity}
\rho_{\phi}=\frac{1}{2}{\dot \phi}^2+V(\phi) \ ,
\end{equation}
\begin{equation}
\label{pressure}
p_{\phi}=\frac{1}{2}{\dot \phi}^2-V(\phi) \ ,
\end{equation}
respectively, with $V(\phi)$ the potential, and where we have defined the Planck mass by $G\equiv M_{\rm p}^{-2}$. The scalar field
itself obeys
\begin{equation}
\label{phieqn}
{\ddot \phi}+3\left(\frac{{\dot a}}{a}\right){\dot \phi}+\frac{dV(\phi)}{d\phi}=0 \ ,
\end{equation}
with a prime denoting a derivative with respect to $\phi$.

Now, to explain the current data indicating an accelerating universe, it is necessary to have the
dominant type of matter at late times be such that ${\ddot a}>0$. If this matter is to be $\phi$, 
then~(\ref{acceleration}) implies $\rho_{\phi} +3p_{\phi} <0$ which, since we conventionally write 
$p_{\phi}\equiv w_{\phi}\rho_{\phi}$ translates into an 
equation of state parameter that obeys
\begin{equation}
\label{phieos}
w_{\phi}<-\frac{1}{3} \ .
\end{equation}
With an appropriate choice of potential the resulting cosmic acceleration can be arranged to occur
at late times, and provides an explanation for the supernova
data~\cite{Riess:1998cb,Perlmutter:1998np}.

Let us now consider extending these ideas to incorporate
inflation~\cite{Spokoiny:1993kt}-\cite{Dimopoulos:2001ix}. This further constrains our potential, and
we will present a general analysis, indicating where appropriate how the particular results 
of~\cite{Peebles:1998qn} are recovered.

Consider a generic potential which, for convenience and for comparison with other work, we will
express in the form
\begin{equation}
V(\phi)=\left\{\begin{array}{lll}
V_1(\phi) & \ \ \ \ \ , \ \ \ \ \ \ & \phi\in (-\infty,0] \\
V_2(\phi) & \ \ \ \ \ , \ \ \ \ \ \ & \phi\in (0,\infty) \ ,
\end{array}\right.
\end{equation}
such that $V_1(\phi)\geq 0$, $V_2(\phi)\geq 0$ $\forall \phi$, $V_1(0)=V_2(0)$ and $V_1'(0)=V_2'(0)$.  We shall in general be interested in cases 
where $(V(\phi)$ is monotonically decreasing and concave ($V'(\phi)<0$ and $V''(\phi)>0$ $\forall \phi$). This is a generalized form of the potential used
by Peebles and Vilenkin~\cite{Peebles:1998qn} in which
\begin{eqnarray}
V_1^{\rm PV}(\phi) & = & \lambda (\phi^4 +M^4) \ , \\
V_2^{\rm PV}(\phi) & = & \frac{\lambda M^8}{\phi^4 +M^4} \ . 
\end{eqnarray}

Since we require inflation to occur at early times,
when the expectation value of $\phi$ is large and negative, we would like $V_1(\phi)$ to satisfy
the slow-roll conditions 
\begin{eqnarray}
\epsilon \equiv \left[\frac{V_1'(\phi)}{V_1(\phi)}\right]^2 \frac{M_{\rm p}^2}{16\pi} & \ll & 1 \nonumber \\
{\bar \eta} \equiv \left|\frac{V_1''(\phi)}{V_1(\phi)}\right| \frac{M_{\rm p}^2}{8\pi} & \ll & 1 \ ,
\end{eqnarray}
for sufficiently large and negative $\phi$.

Inflation ends when the slow-roll conditions are violated and the potential and
kinetic energies of the inflaton are comparable with each other. We denote this epoch by the
subscript $x$, following Peebles and Vilenkin~\cite{Peebles:1998qn}, so that the above statement reads
\begin{equation}
V_1(\phi_x)\simeq \frac{1}{2}{\dot \phi}_x^2 \ .
\label{potkinequal}
\end{equation}
In the models in this paper we will always have $\phi_x \simeq -M_{\rm p}$ and~(\ref{friedmann}) then
implies that the Hubble parameter at this epoch is given by
\begin{equation}
\label{hubbleendinflation}
H_x=\sqrt{\frac{8\pi V_1(-M_{\rm p})}{3M_{\rm p}^2}} \ .
\end{equation}

In traditional inflationary models the inflaton then rapidly transfers its energy to other fields either through perturbative effects (reheating) or parametric resonance (preheating). Here, however,
there is no such effect, and it is the kinetic energy of the field $\phi$ that is the dominant component
of the energy density of the universe immediately after the end of inflation. Following 
Joyce~\cite{Joyce:1996cp} we term this behavior {\it kination}.
Since time derivatives
of scalar fields scale as ${\dot \phi}(t)\propto a(t)^{-3}$ we may use~(\ref{potkinequal}) to show that the
evolution of the energy density in $\phi$ during the kination era obeys
\begin{equation}
\rho_{\phi}(a)\simeq V_1(-M_{\rm p})\left(\frac{a_x}{a}\right)^6 \ .
\end{equation}

Now, assuming spatial flatness (which the epoch of inflation will ensure in general), we may solve
the Friedmann equation~(\ref{friedmann}) for the cosmic scale factor to obtain
\begin{equation}
\left[\frac{a(t)}{a_x}\right]^3=3\sqrt{\frac{8\pi V_1(-M_{\rm p})}{3M_{\rm p}^2}} t \ .
\end{equation}
Since the universe is kinetic energy dominated during this epoch, we may also obtain
\begin{equation}
\phi(t) = \frac{M_{\rm p}}{\sqrt{12\pi}} \ln\left(\frac{t}{t_x}\right)-M_{\rm p} 
= \frac{3M_{\rm p}}{\sqrt{12\pi}} \ln\left(\frac{a}{a_x}\right)-M_{\rm p} \ ,
\end{equation}
where we have imposed $\phi(t_x)=-M_{\rm p}$ and
\begin{equation}
t_x = \frac{1}{3}\sqrt{\frac{3M_{\rm p}^2}{8\pi V_1(-M_{\rm p})}}
\end{equation}
is the cosmic time at which inflation ends.

A successful cosmology requires the universe be radiation-dominated at the time of nucleosynthesis,
since otherwise the precision predictions of that theory are no longer in agreement with
observations. The lack of conventional reheating, the conversion of the potential energy of the
inflaton to particle production, in quintessential inflation means that the requisite radiation must
be produced another way. In fact, the radiation era in these models is due to the subtle behavior
of quantum fields in changing geometries.

At the end of inflation, the FRW line element undergoes an abrupt change from that associated with
cosmic expansion (exponential or power-law) to that associated with kination. Massless quantum
fields in their vacua in the inflation era are no longer in vacuum in the kination era, corresponding
to gravitational particle production. This effect is analogous to Hawking radiation, and has been explored
in detail~\cite{Ford:1986sy}-\cite{Giovannini:1998bp} in the cosmological context of interest here.

The radiation density produced in this way is
\begin{equation}
\label{radiationdensity}
\rho_r = RH_x^4 \left(\frac{a_x}{a}\right)^4 \ ,
\end{equation}
where $R\sim 10^{-2}$, and the number density of massless particles produced is given by
\begin{equation}
\label{radnumberdensity}
n\sim RH_x^3 \left(\frac{a_x}{a}\right)^3 \ .
\end{equation}
At such early times in the universe, thermal equilibrium is not yet established due to the rapid pace
of cosmic expansion. The massless particles produced by the effects of quantum fields in our changing
space-time only establish thermal equilibrium when the Hubble parameter has dropped to a value
$n\sigma \sim H$, where $\sigma$ is the particle-antiparticle annihilation cross-section.
 
Since
\begin{equation}
\label{crosssection}
\sigma \sim \frac{\alpha^2 a^2}{H_x^2 a_x^2} \ ,
\end{equation}
where $\alpha \sim 0.01-0.1$ is a coupling constant, we obtain
\begin{equation}
\label{athoverax}
\frac{a_{\rm th}}{a_x} \sim \frac{1}{\alpha\sqrt{R}} \ .
\end{equation}
Therefore, thermalization takes place at a temperature
\begin{equation}
\label{ }
T_{\rm th}\sim \alpha R^{3/4} \sqrt{\frac{8\pi V_1(-M_{\rm p})}{3M_{\rm p}^2}} \ .
\end{equation}
This is the highest temperature at which there is thermal equilibrium in the universe.
 
Now that we know how both the scalar field and the radiation evolve during the kinetic energy
dominated era, we may easily calculate the scale factor at which radiation-domination occurs.
Demanding that $\rho_{\phi}(a_r)=\rho_r(a_r)$ we obtain
\begin{equation}
\label{ }
\frac{a_r}{a_x}\sim \frac{3M_{\rm p}^2}{8\pi \sqrt{RV_1(-M_{\rm p})}} \ ,
\end{equation}
and the temperature $T_r$ at which radiation domination begins is then simply calculated to be
\begin{equation}
\label{TR}
T_r \sim \left(\frac{8\pi}{3}\right)^{3/2} R^{3/4}\frac{V_1(-M_{\rm p})}{M_{\rm p}^3} \ ,
\end{equation}
so that we have
\begin{equation}
\label{TthoverTr }
\frac{T_{\rm th}}{T_r} \sim \left(\frac{3}{8\pi}\right) \alpha \frac{M_{\rm p}^2}{\sqrt{V_1(-M_{\rm p})}} \ .
\end{equation}

Thus, for $\alpha\sim 0.1$, since the scale of the potential is no greater than $M_{\rm p}^4$, we
see that $T_{\rm th}>T_r$ and so when radiation-domination begins the universe is immediately
in thermal equilibrium.

Clearly, to obtain our standard cosmology it is necessary to have a period of radiation domination
(followed by matter domination) before dark energy domination begins. This means requiring 
that $a_r <a_*$, so that at $T_r$ the universe becomes dominated by radiation, with the scalar field evolving in the
background, its potential and kinetic energies subdominant to the radiation. 
In particular, nucleosynthesis takes place at $T_{\rm nuc}\sim 1$MeV. To ensure that the universe is
radiation dominated at this epoch we should conservatively require $T_r\geq 10$MeV. 
Using~(\ref{TR}), this allows
us to bound the energy scale associated with quintessential inflation by
\begin{equation}
\label{ }
V_1^{1/4}(-M_{\rm P})\geq 10^{-5}M_{\rm P} \ .
\end{equation}
Such a bound does not exist for the standard inflationary paradigm because reheating effects
ensure that the universe is radiation-dominated immediately following inflation. In quintessential
inflation however, there is a comparatively small amount of radiation produced through gravitational
particle production, so that radiation-domination occurs much later.

For a significant 
time subsequent to this, cosmic evolution is much the same as in the standard cosmology, with
a matter dominated epoch eventually succeeding the radiation era. Although the scalar field is not
important during these times, the density fluctuations seeded by quantum fluctuations in $\phi$ 
during inflation lead to structure formation and temperature fluctuations in the cosmic microwave 
background radiation. In order to obtain agreement with the COBE anisotropy measurements, we
must require
\begin{equation}
\label{flucts}
\frac{V^{3/2}(\phi_i)}{M_{\rm p}^3|V'(\phi_i)|} \sim 5.2\times 10^{-5} \ ,
\end{equation}
where $\phi_i$ denotes the value of $\phi$ 60 efolds before the end of inflation. Note that in the
Peebles-Vilenkin model this constraint translates to $\lambda\sim 10^{-14}$, similar to that tuning 
required by standard chaotic inflationary potentials.

Finally, let us turn to this extreme future of the universe, in which the scalar field is rolling in the
potential $V_2(\phi)$ and becomes responsible for quintessence at an epoch denoted by a subscript
$*$. It is clear that when quintessence begins, the energy density of the field $\phi$ once again
becomes dominated by its potential energy density. This implies
\begin{equation}
V_2(\phi_*)=V_1(-M_{\rm p})\left(\frac{a_x}{a_*}\right)^6 \ ,
\end{equation}
where
\begin{equation}
\phi_* =\frac{3M_{\rm p}}{\sqrt{12\pi}}\ln\left(\frac{a_*}{a_x}\right) -M_{\rm p} \ .
\end{equation}

It is a challenge similar to that for conventional quintessence to ensure that this epoch occurs at
the present time and yields the correct ratio of matter to dark energy. However, such considerations
apply far after baryogenesis and we shall refer the reader to other treatments for the details of how
this occurs~\cite{Spokoiny:1993kt}-\cite{Dimopoulos:2001ix}.

\section{Quintessential Baryogenesis}
\label{qbg}
In order for the quintessence field $\phi$ to play a role in baryogenesis, we must consider how $\phi$
couples to other fields. In principle, the inflaton and quintessence field may lie in any sector of the 
theory, the phenomenologically safest of which would be one in which there are only gravitational 
strength
couplings to other particles. This is presumably the best we can do, since attempts to protect 
gravitational-strength couplings through global symmetries can be thwarted by wormholes and quantum
gravitational effects~\cite{Giddings:cg}-\cite{Kallosh:1995hi}. We will adopt a conservative approach
and assume that $\phi$ couples to standard model fields with couplings specified by a dimensionless
constant and an energy scale which we shall leave as a free parameter for the moment and later 
constrain by observations and the condition that our model produce a sufficient baryon asymmetry.

We consider terms in the effective Lagrangian density of the form
\begin{equation}
\label{jmucoupling}
{\cal L}_{\rm eff}=\frac{\lambda'}{M}\partial_{\mu}\phi J^{\mu} \ ,
\end{equation}
where $\lambda'$ is a coupling constant, $M<M_{\rm p}$ is the scale of the cutoff in the effective theory
and $J^{\mu}$ is the current corresponding to some continuous global symmetry such as baryon
number or baryon number minus lepton number. Further, let us
assume that $\phi$ is homogeneous. We then obtain
\begin{equation}
\label{coupling}
{\cal L}_{\rm eff}=\frac{\lambda'}{M}{\dot \phi}\ \Delta n \equiv \mu(t)\Delta n \ ,
\end{equation}
where $n=J^0$ is the number density corresponding to the global symmetry and we have
defined an effective time-dependent ``chemical potential'' $\mu(t)\equiv \lambda'{\dot \phi}/M$. 

Recall that we need to satisfy the Sakharov criteria in order to generate a baryon asymmetry 
(for reviews see~\cite{Cohen:1993nk}-\cite{Riotto:1999yt}). 
The first of these requires baryon number $B$ to be violated. At this stage, to maintain 
generality, we shall leave the mechanism of baryon number violation unspecified, and will address
particular cases later. Possible sources are the decay of superheavy grand-unified gauge bosons or
anomalous electroweak processes at finite temperature.
Further, the standard
model is maximally C-violating due to its chiral structure, and the coupling~(\ref{jmucoupling}) is
$CP$-odd. In this sense, no {\it explicit} $CP$-violation is required in this model.
The third Sakharov criterion requires a departure from thermal equilibrium if $CPT$ is a manifest symmetry. However, the crucial point about baryogenesis in the presence of the rolling scalar
field $\phi$ is that $CPT$ 
is broken spontaneously by the explicit value taken by $\langle{\dot \phi}\rangle \neq 0$.
Thus, the particular model of baryogenesis that is important here is 
{\it spontaneous baryogenesis}~\cite{Cohen:1988kt},
which is effective even in thermal equilibrium.
We will refer to this model, in which the rolling scalar responsible for both inflation and dark energy
also provides a source for spontaneous baryogenesis as {\it quintessential baryogenesis}. In our
model it is the field $\phi$ that plays the role of Cohen and Kaplan's {\it thermion}~\cite{Cohen:1988kt}. The idea that the inflaton could drive spontaneous baryogenesis was discussed briefly 
in~\cite{Cohen:1988kt}, where it was correctly noted that accelerated expansion would reduce the
baryon number generated {\it during} inflation to a negligible magnitude, and therefore that barogenesis during the
reheating phase was the only possibility. However, as we shall see, in the context of quintessential
inflation there exists a significant range of postinflationary cosmic history during which
spontaneous baryogenesis may occur.

To understand how spontaneous (and hence quintessential) baryogenesis works, note that in thermal equilibrium we have
\begin{equation}
\Delta n(T;\xi)=\int \frac{d^3{\bf p}}{(2\pi)^3}[f(E,\mu)-f(E,-\mu)] \ ,
\end{equation}
where $\xi\equiv \mu/T$ is a parameter and $f(E,\mu)$ is the phase-space distribution of the particles
of the current $J^{\mu}$, which may be Fermi-Dirac or Bose-Einstein. Thus, for $\xi <1$
\begin{equation}
\Delta n(T;\mu)\simeq \frac{gT^3}{6} \xi+{\cal O}(\xi^2) \ ,
\label{deltan}
\end{equation}
where $g$ is the number of degrees of freedom of the field corresponding to $n$. Therefore,
\begin{equation}
\Delta n(T;\mu)\simeq \frac{\lambda' g}{6M} T^2{\dot \phi} \ .
\end{equation}

Now recall that the entropy density is given by
\begin{equation}
s=\frac{2\pi}{45} g_* T^3 \ ,
\end{equation}
where $g_*$ is the effective number of relativistic degrees of freedom in thermal equilibrium at
temperature $T$. Whatever the mechanism of baryon number violation, there will exist a temperature
$T_F$ below which baryon number violating processes due to this mechanism become sufficiently
rare that they freeze out. For $T<T_F$ these processes can no longer appreciably change the baryon 
number of the universe. Computing the freeze-out value of the baryon to entropy ratio we then obtain
\begin{equation}
\label{baueqn}
\eta_F\equiv \eta(T_F) \equiv \frac{\Delta n}{s}(T_F) \simeq 0.38 \lambda' \left(\frac{g}{g_*}\right)\frac{{\dot \phi}(T_F)}{MT_F} \ .
\end{equation}

Quintessential baryogenesis is effective at temperatures $T_{\rm th} >T>T_F$, with corresponding
scale factors $a_{\rm th}\equiv a(t_{\rm th})<a(t)<a_F\equiv a(t_F)$. We have seen that 
$T_{\rm th}<T_r$, so that baryon number violating interactions are first in equilibrium 
during the kination epoch in which the evolution of the scalar field can be written as
\begin{equation}
{\dot \phi}\simeq \sqrt{2V_1(-M_{\rm p})} \left(\frac{a_x}{a}\right)^3 \ ,
\end{equation}

If we assume that $T_F >T_r$ then~(\ref{baueqn}) yields
\begin{equation}
\eta_F \sim 3.8\times 10^{-3}\sqrt{\frac{2 V_1(-M_{\rm p})}{M_{\rm p}^4}}
\frac{M_{\rm p}}{T_F}\left(\frac{M_{\rm p}}{M}\right)
\left(\frac{a_x}{a_F}\right)^3 \lambda' \ ,
\end{equation}
where we have used $(g/g_*)\sim 10^{-2}$.
Writing, for convenience,
\begin{equation}
\label{ }
a_F=\gamma a_{\rm th} \ ,
\end{equation}
with $1\leq \gamma \leq 10^6$ (so that $T_F=T_{\rm th}/\gamma$) and using~(\ref{athoverax}) 
we then obtain
\begin{equation}
\label{ }
\eta_F \sim 5.4\times 10^{-3}\times \sqrt{\frac{8\pi}{3}} \lambda' \gamma^{-2} R^{3/4} \alpha^2
\left(\frac{M_{\rm p}}{M}\right) \ .
\end{equation}
Inserting the values for $R$ and $\alpha$ yields
\begin{equation}
\label{etaFfinal}
\eta_F \simeq 2 \times 10^{-7}\times \lambda' \gamma^{-2} \left(\frac{M_{\rm p}}{M}\right) \ .
\end{equation}
Notice that, as expected, this final result is linear in the effective chemical potential, and contains
a power of $\gamma$ reflecting the appropriate amount of redshifting occurring during the 
kination epoch between the temperatures $T_{\rm th}$ and $T_F$.

To make further progress we must calculate $T_F$ and $\phi(T_F)$ in order to find $\gamma$, and
this requires knowledge of the dynamics of $\phi$. To do this correctly
it is necessary to consider the possible effects of back-reaction of the coupling to $J^{\mu}$ on the
dynamics of $\phi$. Taking account of the effective Lagrangian into account, the equation of
motion of $\phi$ becomes
\begin{equation}
{\ddot \phi}+3H{\dot \phi} +V'(\phi)+\frac{\lambda'}{M}{\dot {\Delta n}}
+3\frac{\lambda'}{M}H\Delta n=0 \ .
\end{equation}
Using~(\ref{deltan}) we obtain
\begin{equation}
\left[1+\frac{\lambda' g}{6}\left(\frac{T}{M}\right)^2\right]\left({\ddot \phi}+3H{\dot \phi}\right)
+V'(\phi)=0 \ .
\end{equation}
Therefore, for $T<M$, we are justified in neglecting the extra term and we may safely
neglect the back-reaction on $\phi$. This approximation will typically easily be satisfied, and in
particular it is well-justified in the Peebles-Vilenkin model~\cite{Peebles:1998qn} in which $T\ll M$.

How the baryon excess evolves after this point depends on the value of $T_F$ and on the
relevant current in equation~(\ref{jmucoupling}). If 
$T_F \leq T_c^{\rm EW}\sim 100$ GeV, the critical temperature of the electroweak phase transition,
then all baryon number violation ceases at $T_F$ and $\eta(T<T_F)=\eta_F$. However, if
$T_F>T_c^{\rm EW}$, then we must take into account the effects of anomalous electroweak
processes at finite temperature. These can be involved in directly generating the baryon asymmetry
(electroweak baryogenesis), in reprocessing an asymmetry in other quantum numbers into one in
baryon number (for example in leptogenesis) or in diluting the asymmetry created by any 
baryogenesis mechanism which is effective above the electroweak scale and does not produce a
$B-L$ asymmetry. It is important to realize that, in the context of quintessential inflation, the
quantitative effects of these electroweak processes may differ substantially from those in the standard 
cosmology~\cite{Joyce:1996cp,Joyce:1997fc} since in our case, 
the electroweak phase transition may occur during kination rather than radiation domination.

In the electroweak theory baryon number violating processes at zero temperature are mediated by a
saddle-point field configuration known as the sphaleron~\cite{Klinkhamer:1984di}. We shall therefore
refer to the finite temperature configurations relevant here as thermal sphalerons.

However, given the constraints we shall present in section~\ref{sec:constraints}, we shall see
that it is necessary to have $\gamma \geq 10^2$ and no electroweak dilution in order to generate a sufficient baryon asymmetry.

We have now provided quite a general description of quintessential baryogenesis. While this has
allowed us to demonstrate the generic features of our model, we cannot calculate the magnitude of
the actual baryon asymmetry generated without first specifying a mechanism of baryon number
violation (and hence a value for $T_F$) and a value for the dimensionless combination 
$\lambda' M_{\rm p}/M$. Let us now turn to some concrete examples.

\subsection{Baryon Number Violation Through Non-renormalizable Operators}
If there exists baryon number violating physics above the standard model, then this physics
will manifest itself in non-renormalizable operators  in the standard model. For the purposes of this
section we will actually be interested in operators that violate the anomaly-free combination
$B-L$. In that case the value $\eta_F$ calculated via ~(\ref{etaFfinal}) will be the final
baryon to entropy ratio $\eta$, since anomalous electroweak processes preserve this combination
of quantum numbers. Consider the effective 4-fermion operator
\begin{equation}
\label{4fermion}
{\cal L}_{B-L}=\frac{{\tilde g}}{M_X^2} \psi_1\psi_2{\bar \psi}_3{\bar \psi}_4 \ ,
\end{equation}
where $\psi_i$ denote standard model fermions. Here ${\tilde g}$ is a dimensionless coupling, obtained after integrating out the $B-L$ violating effects of a particle of mass $M_X$. 
The rate of baryon number violating processes due to this operator is, as usual, defined by
$\Gamma_{B-L}(T)=\langle \sigma(T) n(T) v \rangle$, where $\sigma (T)$ is the cross-section for
$\psi_1 + \psi_2 \rightarrow \psi_3 + \psi_4$, $n(T)$ is the number density of $\psi$ particles, $v$
is the relative velocity and $\langle \cdots \rangle$ denotes a thermal average. 
For temperatures
$T<M_X$ we have $n(T)\sim T^3$, $\sigma (T) \sim {\tilde g}^2 T^2/M_X^4$, and $v\sim 1$ which 
yields
\begin{equation}
\label{ }
\Gamma_{B-L}(T)\simeq \frac{{\tilde g}^2}{M_X^4} T^5 \ .
\end{equation}
The high power of the temperature dependence in this rate results from the fact 
that~(\ref{4fermion}) is an irrelevant operator in the electroweak theory and, as we shall see, is
crucial for the success of our mechanism.
These interactions are in thermal equilibrium in the early universe, but because their rate drops
off so quickly with the cosmic expansion they will drop out of equilibrium at the temperature $T_F$ 
defined through
\begin{equation}
\label{ }
\Gamma_{B-L}(T_F)=H(T_F) \ .
\end{equation}
Thus,
\begin{equation}
\label{ }
T_F \simeq \left(\frac{3}{8\pi}\right)^{3/4}\frac{1}{{\tilde g}R^{3/8}}
\sqrt{\frac{M_{\rm p}^2}{V_1(-M_{\rm p})}} M_X^2 \ .
\end{equation}
As a definite example, let us take $\lambda' M_{\rm p}/M\sim 8$. As we shall see in~\ref{sec:constraints},
this is as large as is allowed by
current constraints. In this case~(\ref{etaFfinal}) implies that we need $\gamma\sim 10^2$ in
order to obtain the correct BAU. This 
then implies that we must have
\begin{equation}
\label{ }
M_X\sim\left(\frac{8\pi}{3}\right)^{5/8}R^{9/16}
\sqrt{\frac{\alpha {\tilde g} V_1(-M_{\rm p})}{M_{\rm p}^2}} \ .
\end{equation}
For reference, note that in the Peebles-Vilenkin model~\cite{Peebles:1998qn} this becomes
$M_X^{\rm PV}\sim 10^{11}$ GeV, the intermediate scale that appears in some supersymmetric 
models.

\subsection{Grand Unified Baryon Number Violation}
A natural source of baryon number violation to consider is that arising from gauge-mediated
interactions in grand unified theories (GUTs), in which quarks and leptons lie in the same representation 
of the GUT gauge group. However, as we shall demonstrate briefly here, this is not a viable source
for quintessential baryogenesis.

Consider a GUT gauge boson $X$ with mass $M_X$. This particle decays through a
renormalizable coupling with decay rate
\begin{equation}
\label{ }
\Gamma_{\rm GUT} \simeq \frac{{\tilde g}_X^2}{16\pi} \left\{\begin{array}{ll}
M_X      &    \ \ \ \ \ \mbox{for $T\leq M_X$}\\
T     &       \ \ \ \ \ \mbox{for $T\geq M_X$}
\end{array} \right. \ .
\end{equation}
It is then relatively straightforward to show that $X$ particles decay in equilibrium (as is required
by our mechanism) only if 
\begin{equation}
\label{ }
M_X \leq 10^9 \left(\frac{T_r}{\rm GeV}\right)^{1/2} \mbox{GeV} \ .
\end{equation}
Since GUT gauge bosons are much heavier than this value this source of baryon number violation is
not useful for our model.

One might be tempted to consider a particle with the appropriate mass produced by some physics 
beyond the standard model. However, it quickly becomes apparent that decays through renormalizable 
interactions can never work, since these particles then decay and the interactions are never again 
out of equilibrium (there is no freeze-out). Thus, as we commented earlier, nonrenormalizable
operators seem essential for decays of massive particles to work.

\subsection{Electroweak Baryon Number Violation}
Another important source of baryon number violation in baryogenesis models comes from anomalous
electroweak processes at finite temperature. Thermal sphalerons are
in equilibrium at temperatures above the critical temperature of the electroweak phase
transition $T_c^{\rm EW}\sim 10^2$GeV and
thus, the final baryon to entropy ratio generated purely through electroweak processes
is given by~(\ref{etaFfinal}) with $\gamma_{\rm EW}\sim 10^7$. This yields
\begin{equation}
\label{etatEWfinal}
\eta^{\rm EW}\sim 2\times 10^{-21}\left(\frac{\lambda' M_{\rm p}}{M}\right) \ ,
\end{equation}
which, since $\lambda' M_{\rm p}/M<8$, is far too low to explain the observed baryon 
asymmetry~(\ref{BAUobserved}).

\section{Constraints and Tests}
\label{sec:constraints}
The presence of an extremely light scalar field in the universe has the potential to lead to a number
of observable consequences in the laboratory and in cosmology. In the case of quintessence these
effects have been analyzed in some detail by Carroll~\cite{Carroll:1998zi}. Particularly strong constraints
arise due to couplings of the form
\begin{equation}
\label{phiFFdual}
{\cal L}_{\rm eff}^{(1)}\equiv \beta_{F{\tilde F}}\frac{\phi}{M}F_{\mu\nu}{\tilde F}^{\mu\nu} \ ,
\end{equation}
where $F_{\mu\nu}$ and 
${\tilde F}^{\mu\nu}\equiv \frac{1}{2}\epsilon^{\mu\nu\rho\sigma}F_{\rho\sigma}$ are the
electromagnetic field strength tensor and its dual respectively. If, as in quintessence models, the field
$\phi$ is homogeneous and time varying, then it affects the dispersion relation for electromagnetic
waves and leads to a rotation in the direction of polarized light from radio 
sources~\cite{Carroll:vb,Carroll:1991zs}. The bound obtained is
\begin{equation}
\label{seanbound}
\left|\beta_{F{\tilde F}}\right| \leq 3\times 10^{-2} \left(\frac{M}{M_{\rm p}}\right) \ ,
\end{equation}
where we have assumed that $\phi$ rolls over about a Planck mass during the last half a redshift of
the universe.

To avoid this bound, quintessence models usually struggle to have such couplings be as small as
possible. In our model however, we are making important use of the coupling~(\ref{jmucoupling}). If
the relevant current is that for baryon number $J^{\mu}_B$, then using the anomaly equation we may
rewrite our term~(\ref{jmucoupling}) as
\begin{equation}
\label{anomaly}
{\cal L}_{\rm eff}=-\frac{\lambda'}{M}\phi
n_f\left(\frac{g^2}{32\pi^2}W_{\mu\nu}^a {\tilde W}^{a\mu\nu}
-\frac{g'^2}{32\pi^2}B_{\mu\nu}{\tilde B}^{\mu\nu}\right) \ ,
\end{equation}
where $g$ and $g'$ are the gauge couplings of $SU(2)_L$ and $U(1)_Y$, 
respectively,  $n_f$ is the number of families and 
$\tilde{W}^{\mu\nu}$ and $\tilde{B}^{\mu\nu}$
are the duals of the $SU(2)_L$ and $U(1)_Y$ field strength tensors respectively.

Thus, the constraint~(\ref{seanbound}) should apply to give
\begin{equation}
\label{ourseanbound}
\left(\frac{n_f g'^2}{32\pi^2}\right) \lambda' \leq 3\times 10^{-2} \left(\frac{M}{M_{\rm p}}\right) \ ,
\end{equation}
Putting in the appropriate standard model numbers we obtain
\begin{equation}
\label{seannumbers}
\lambda' \frac{M_{\rm p}}{M} < 8 \ .
\end{equation}

If the relevant current is $J^{\mu}_{B-L}$ then the above argument does not apply formally.
However, we may still generally expect an analogous coupling to the term 
$F^{\mu\nu} {\tilde F}_{\mu\nu}$ of a similar order.

We have already demonstrated that, for an appropriate scale $M_X$, 
successful quintessential baryogenesis takes place for 
$\lambda'  < 8$, and so it is possible to generate the observed BAU and to evade existing
constraints. If quintessential baryogenesis is correct, then it may be that the relevant coupling
$\lambda'$ lies just below the existing observational bounds and that future studies of the
rotation of polarized light from distant galaxies will reveal the presence of such a term.

One might also worry about the laboratory constraints on the operator corresponding to a coupling
to $F^{\mu\nu} {\tilde F}_{\mu\nu}$ resulting from~(\ref{anomaly}) after electroweak
symmetry breaking.
For example, in the standard model these can lead to contributions to the electric dipole moments
of the electron and the neutron. These effects have been considered~\cite{Zhang:1993vh,Lue:1996pr} 
in the context of electroweak baryogenesis. Applying the experimental 
bounds~\cite{Commins:gv,smithaltarev} the relevant bound is significantly weaker than the cosmological
one quoted above.

We should also comment briefly on the naturalness of the potentials required. We
have taken the quintessence and quintessential inflation paradigms and explored another 
generic effect in these models. It should be pointed out however, that the potentials require a great
deal of fine-tuning. In particular, the extremely small value of the self-coupling $\lambda$ to explain the
fluctuations in the CMB is unexplained. If we consider embedding the model in a supersymmetric theory
we will come up against the problems that standard quintessence faces, that the flatness of the potential
required at late times in the universe may be ruined by the soft supersymmetry breaking terms 
required to obtain a realistic phenomenology today.

\section{Comments and Conclusions}
\label{conclusions}
Modern particle cosmology concerns the search for dynamical explanations for the initial conditions
required by the standard FRW cosmology. Particular attention has been paid recently to one of these
initial condition problems, that of the size of the vacuum energy contribution to the total energy density
of the universe. The root of this issue is that vacuum energy, or something approximating it, can
lead to the acceleration of the universe.

The best fit cosmology to all current observational data is one in which the universe undergoes two
separate epochs of acceleration. The first of these, inflation, is the only clear way to seed adiabatic,
scale-free perturbations in the cosmic microwave background radiation. The second epoch, that of
dark-energy domination, is required to simultaneously understand the power spectrum of the CMB
and the expansion of the universe at intermediate redshifts as revealed by type IA supernovae.
These requirements have led cosmologists to introduce a new scalar field to account for the newly
required late-time acceleration.

In this paper we have extended the approach of Spokoiny~\cite{Spokoiny:1993kt}
and of Peebles and Vilenkin~\cite{Peebles:1998qn} in exploring the extent to which the dynamics
of a single scalar field can be responsible for setting multiple parameters required by the standard
cosmology. Our contribution has been to generalize the mechanisms by which inflation and dark
energy domination may be due to a single scalar and to introduce the idea that this same rolling
scalar might be responsible for generating the baryon asymmetry of the universe. The mechanism
that we propose, quintessential baryogenesis, is an application of the spontaneous baryogenesis
model of Cohen and Kaplan\cite{Cohen:1988kt} to the quintessential inflation case.

It seems a particularly powerful idea to us that a single rolling scalar field might be responsible
for a number of the fundamental initial conditions required to make the standard cosmology
work. In the case of the baryon asymmetry, this allows us to associate the existence of an asymmetry 
with the spontaneous breaking of $CPT$ and the direction of the rolling of the
scalar field. 
\begin{figure}
\epsfig{file=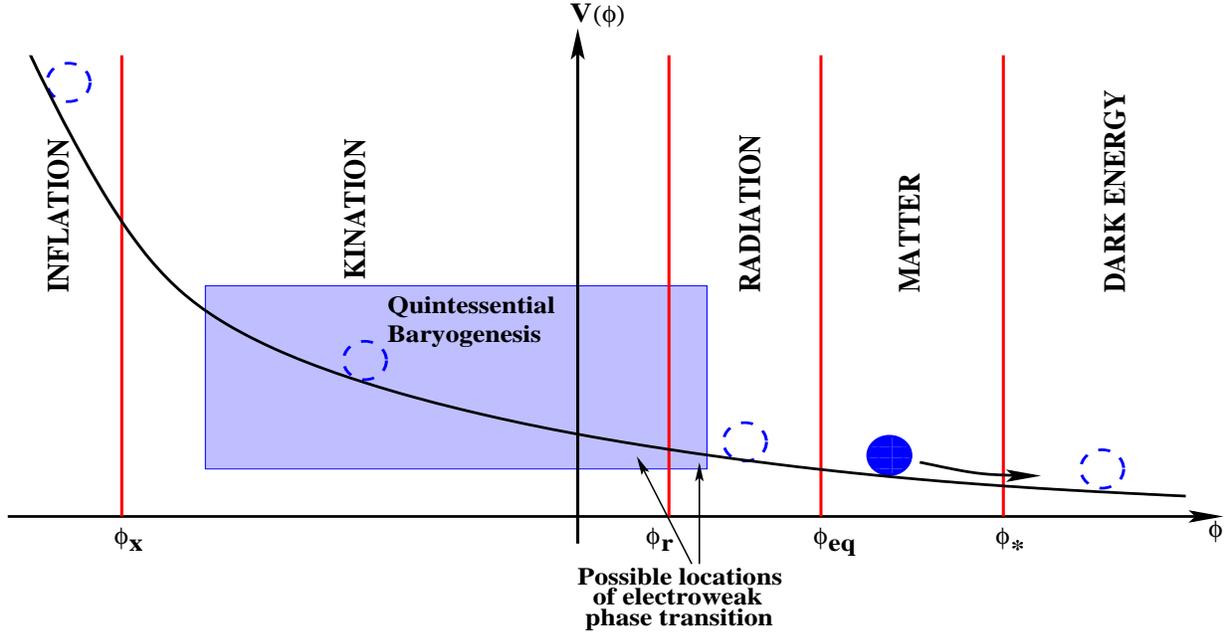, height=3.3in, width=6.4in}
\caption{The evolution of the universe in this model. As the scalar field rolls down its potential
the universe goes through a succession of phases, beginning with inflation, 
generating the baryon asymmetry along the
way, and ending with dark energy domination.}
\label{fig:graphic}
\end{figure}

The evolution of the universe we envisage may be summarized as follows (see figure~\ref{fig:graphic}). 
At the earliest times in the
universe, inflation occurs due to the potential energy dominance of the field $\phi$ which begins
rolling at very large and negative values. Inflation ends when the kinetic energy of the scalar field
becomes important and the slow-roll conditions are violated. Since our potential does not have
a minimum at finite $\phi$, unlike typical inflationary models, conventional reheating does not occur.
Instead, matter is created gravitationally due to the mismatch of vacuum states between the 
approximately de-Sitter state of inflation and that of the kination era. Since kinetic energy density
redshifts more rapidly that radiation energy density, the universe eventually becomes 
radiation-dominated. At this stage the rolling scalar has negligible effect on the expansion rate
of the universe. However, the direction of rolling spontaneously violates $CPT$. If $\phi$ couples to
other fields, as we expect it to generically, then the expectation value of the
baryon number operator in this background in thermal equilibrium is nonzero. Thus a baryon excess is 
generated. After the electroweak phase transition baryon number violation is no longer
effective in the universe and the baryon number existing at that time is frozen in. The scalar field
continues to evolve and in the late universe, after matter-domination has begun, its potential energy
can once again become dominant leading to a new period of dark-energy domination.

The couplings required to make quintessential baryogenesis effective may be generated in a number 
of different ways, for example by gravitational effects coupling the inflaton/dark energy
sector to visible sector fields. We have considered the current experimental constraints on the 
necessary operator and have found that there exist considerable regions of parameter space in which 
our mechanism is consistent. Further, it is possible that a restricted region of this parameter space
may be accessible to future experiments.

We have left a number of questions unanswered and will return to them in future work. Perhaps the
most pressing issue is one that plagues rolling scalar models of dark energy in general, namely the
question of technical naturalness of the potentials involved, and their stability to quantum
corrections. We have omitted any discussion of this here, while laying out the general features of
the model, but these issues must be addressed to put our mechanism on firmer ground. For example, it may be most natural to identify the field $\phi$ with a 
pseudo-Goldstone boson~\cite{Dolgov:1996qq,Frieman:1995pm}, since its coupling to 
the current $J^{\mu}$ is derivative. However, this is a general issue for quintessence models, and is
not specific to our baryogenesis mechanism. We have therefore chosen to concern ourselves with this issue separately.

Taken at face value, current observations imply that our universe is entering an accelerating
phase that may be governed by the rolling of a scalar field. If we are to understand the physics of such
a field then it is important that we investigate other ways in which it may impact cosmology and
particle physics. In particular, if, as we have suggested here, the field is responsible for the generation of
the baryon asymmetry, then the result will be a more economical and attractive
cosmology.

\section*{Acknowledgements}
We thank Robert Brandenberger and Alex Vilenkin for helpful comments on a first draft of this paper.
The work of ADF and MT is supported by the National Science
Foundation (NSF) under grant PHY-0094122. SN is supported by the
US Department of Energy (DOE).

\end{document}